\documentclass[conference]{IEEEtran}
\IEEEoverridecommandlockouts

\usepackage{cite}
\usepackage{amsmath,amssymb,amsfonts}
\usepackage{multirow,multicol}
\usepackage{makecell}
\usepackage{color,soul}
\usepackage{hyperref}

\usepackage[table]{xcolor}  
\usepackage{multirow}

\usepackage{graphicx}
\usepackage{textcomp}
\usepackage{xcolor}

\usepackage{threeparttable}

\usepackage{booktabs} 
\usepackage{upgreek}

\newcommand{\MYcomment}[1]{}
\renewcommand{\figurename}{Fig.}

\newcommand{\MYnote}[1]{}

\newcounter{MYtablecntr}
\addtocounter{MYtablecntr}{1}

\newcommand{\MYlabel}{\small {$\bullet$}}

\newcounter{MYenumctrtwo}

\newcounter{MYenumctr}

\usepackage{textcomp}
\usepackage{tabularx}
\usepackage{tikz}

\usepackage{algorithm}
\usepackage{algpseudocode}
\usepackage{xcolor}

\begin{document}

\title{Edge-Native, Behavior-Adaptive Drone System \\ for Wildlife Monitoring}


\author{\IEEEauthorblockN{1\textsuperscript{st} Jenna Kline}
\IEEEauthorblockA{
\textit{The Ohio State University}\\
0009-0006-7301-5774}
\and
\IEEEauthorblockN{2\textsuperscript{nd} Rugved Katole}
\IEEEauthorblockA{
\textit{The Ohio State University}\\
0009-0003-9700-6598}
\and
\IEEEauthorblockN{3\textsuperscript{rd} Tanya Berger-Wolf}
\IEEEauthorblockA{
\textit{The Ohio State University}\\
0000-0001-7610-1412}
\and
\IEEEauthorblockN{4\textsuperscript{th} Christopher Stewart}
\IEEEauthorblockA{
\textit{The Ohio State University}\\
0000-0002-2860-7889}
}

\maketitle

\begin{abstract}
Wildlife monitoring with drones must balance competing demands: approaching close enough to capture behaviorally-relevant video while avoiding stress responses that compromise animal welfare and data validity. Human operators face a fundamental attentional bottleneck: they cannot simultaneously control drone operations and monitor vigilance states across entire animal groups. By the time elevated vigilance becomes obvious, an adverse flee response by the animals may be unavoidable. To solve this challenge, we present an edge-native, behavior-adaptive drone system for wildlife monitoring. This configurable decision-support system augments operator expertise with automated group-level vigilance monitoring. Our system continuously tracks individual behaviors using YOLOv11m detection and YOLO-Behavior classification, aggregates vigilance states into a real-time group stress metric, and provides graduated alerts (alert vigilance → flee response) with operator-tunable thresholds for context-specific calibration. We derive service-level objectives (SLOs) from video frame rates and behavioral dynamics: to monitor 30fps video streams in real-time, our system must complete detection and classification within 33ms per frame. Our edge-native pipeline achieves 23.8ms total inference on GPU-accelerated hardware, meeting this constraint with a substantial margin. Retrospective analysis of seven wildlife monitoring missions demonstrates detection capability and quantifies the cost of reactive control: manual piloting results in 14 seconds average adverse behavior duration with 71.9\% usable frames. Our analysis reveals operators could have received actionable alerts 51s before animals fled in 57\% of missions. Simulating 5-second operator intervention yields a projected performance of 82.8\% usable frames with 1-second adverse behavior duration, a 93\% reduction compared to manual piloting. 

\end{abstract}


\section{Introduction}
\label{intro}

Drones have revolutionized wildlife monitoring by enabling non-invasive observation of animal behavior in natural habitats \cite{kline2025studying, pedrazzi2025advancing, schad2023opportunities}. However, successful drone-based wildlife studies require navigating a fundamental trade-off: drones must approach close enough to capture behaviorally-relevant video, but not so close that they induce stress responses that compromise both animal welfare and data validity. The vigilance escalation problem poses challenges for drone monitoring of wildlife. 
Prey animals exhibit natural vigilance behaviors: periodic scanning for predators that represents normal, unstressed behavior, referred to as chronic vigilance \cite{kline2025kabrtools}. As drones approach, this can escalate: first to alert vigilance (more frequent head-up postures, alert stances), then to flight response where the entire herd flees \cite{schad2023opportunities}. This escalation occurs on the timescale of seconds to tens of seconds, and once flight response begins, the mission has failed: the animals are stressed, their behavior is no longer natural, and collected video is scientifically compromised \cite{schad2023opportunities, cai2025measuringminimizingdisturbancemarine}.

\begin{figure}
    \centering
    \includegraphics[width=0.8\linewidth]{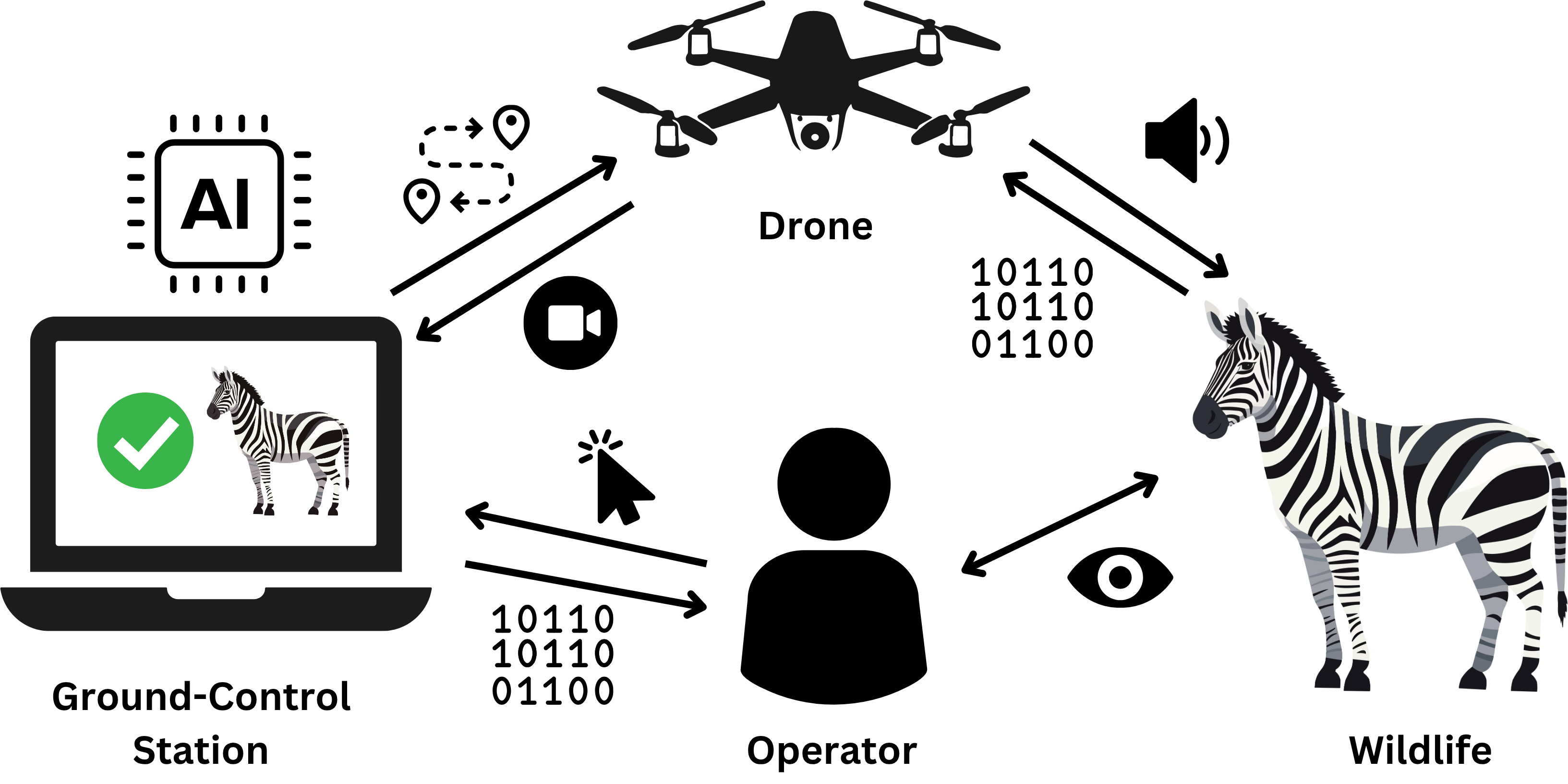}
    \caption{Edge-Native Behavior-Adaptive Drone Control for Wildlife Monitoring. The semi-autonomous drone system augments the operator's ability to observe the wildlife and react in time to prevent adverse behavior events. The ground-control station (GCS) provides the operator with information, allowing them to adjust the automatic, behavior-adaptive navigation appropriately. The GCS recieves video stream from the drone, and in turn sends commands for where the drone should fly next, using a group-tracking navigation policy \cite{kline2025wildwing}, the wildlife monitor the human operator and the drone, and may flee if threatened.}
    \label{fig:teaser}
\end{figure}

Manual (human-in-the-loop, HiTL) piloting faces a fundamental attentional bottleneck: a single operator cannot simultaneously control the drone, maintain safe flight operations, and continuously monitor vigilance states across many individual animals in a herd. By the time elevated vigilance becomes obvious enough for the pilot to notice, it may be too late to prevent flight response. This leads to mission failures that waste expensive field time and risk animal welfare. Our analysis of existing wildlife monitoring datasets reveals that HITL approaches result in 14 seconds average duration of adverse flight behavior per mission, with only 71.9\% of collected frames usable for behavioral analysis. 

Some drone missions, such as mapping large areas for habitat analysis or crop monitoring, are easily automated for consistent, reliable data collection \cite{zhang2020whole, boubin2019managing}.
However, wildlife missions are not easily automated. Human ecological expertise is critical for successful drone missions, making fully autonomous approaches suboptimal. This is because the ecological challenges in the field extends beyond detection to behavior monitoring, and vigilance baselines vary dramatically by context \cite{afridi2025impact}. Habituated animal populations near safari parks or densely populated areas tolerate closer approaches than naive populations; mothers with young exhibit elevated baseline vigilance; habitat structure, time of day, and recent disturbances all affect stress thresholds \cite{schad2023opportunities}. No fixed autonomous control policy can safely handle this variability, thus expert judgment based on species knowledge and field observations remains essential.

We present an edge-native, behavior-adaptive
drone system for terrestrial wildlife monitoring. This is a configurable decision-support system that augments human wildlife monitoring expertise with automated vigilance detection. Rather than replacing the operator, our system acts as an additional pair of eyes: computer vision models continuously monitor all individuals in the herd, quantify vigilance levels in real-time, and alert the operator before elevated vigilance escalates to unrecoverable flight response. The operator retains full control and can tune detection thresholds based on species, population, and field conditions. Just as collision avoidance systems provide preemptive warnings before impact is inevitable, our system monitors leading indicators of animal stress, i.e., elevated vigilance behaviors, allowing operators to implement graduated responses, such as reduced approach speed, pause, or retreat, before a flight response occurs. Autonomous robotics systems for measuring and minimizing disturbance to animals have been proposed in the marine domain \cite{cai2025measuringminimizingdisturbancemarine}. We leverage recent advancements machine learning models capable of quickly inferring wildlife behavior to automatically detect and prevent disturbance in terrestrial wildlife \cite{kholiavchenko2024kabr, chan2025yolo}.


We derive quantitative latency requirements from animal behavior dynamics. Real-time tracking of herd with drones has a suggested service-level-objective (SLO) of 1000ms to keep animals in frame \cite{kline2024characterizing}. However, behavior transitions occur almost instantly, requiring real-time monitoring of the 30 frame per second (fps) video stream, or 33~ms latency \cite{kline2025kabrtools, duporge2025baboonland}. The average time from initial stress response to full flight behavior in zebra herds from seven wildlife missions was found to be 50s. However, as the duration of acute vigilance increases without intervention, so does the likelihood of permanently spooking the animals, ending the mission prematurely and spoiling the behavioral data \cite{schad2023opportunities, pedrazzi2025advancing}. To provide actionable warnings to operators, our system must complete detection and classification within the 33~ms. To reduce false positives, we employ a rolling 3-sample window, similar the strategy employed in Kenyan Animal Behavior Recognition study \cite{kholiavchenko2024kabr, kline2025kabrtools}, so warnings are not triggered erroneously. We demonstrate that our edge-native pipeline, which combines YOLOv11m detection (4.7~ms) with YOLO-Behavior classification (19.1~ms), achieves 23.8~ms total latency on GPU-accelerated edge devices, meeting this requirement with substantial margin.

We evaluate our system using seven wildlife monitoring missions from established wildlife drone datasets (KABR \cite{kholiavchenko2024kabr}, WildWing \cite{kline2025wildwing}, and MMLA \cite{kline2025mmla}) spanning zebras, giraffes, and wild horses across diverse habitats. While retrospective analysis cannot validate closed-loop control performance, it establishes three critical system capabilities: (1) detection models achieve sufficient accuracy on real wildlife footage, (2) inference latency meets behavioral timescale requirements, and (3) the cost of reactive-only control justifies the complexity of behavior-aware systems. 

Our analysis reveals that by detecting elevated vigilance before flight response, our system could provide operators with actionable warning windows. In the future, this warning can be integrated into the control loop to pre-preemptively intervene, i.e. `apply the breaks' to slow the drone automatically. When we simulate operator intervention within 5 seconds of alerts, projected performance shows 82.8\% usable frame yield with only 1 second average adverse behavior duration, a 93\% reduction compared to baseline HiTL approaches.

Our contributions are as follows:

\begin{enumerate}
    \item \textbf{Behavioral timescale-driven SLOs:} We establish quantitative service-level objectives for wildlife monitoring (22s warning window, 25ms control-loop latency) derived from animal behavior dynamics.
    
    \item \textbf{Context-tunable vigilance monitoring:} We present a vigilance scoring algorithm with user-adjustable thresholds that allows ecologists to calibrate the system for species-specific, population-specific, and context-specific baselines.
    
    \item \textbf{Retrospective validation methodology:} We demonstrate detection capability and latency compliance through analysis of real wildlife missions, quantifying warning windows and comparing against reactive control baselines. Our analysis reveals warning windows of 22-91 seconds (mean: 51s) between initial 
vigilance detection and flight response, showing real-time monitoring 
provides actionable intervention.

\end{enumerate}

The remainder of this paper is organized as follows: Section \ref{background} reviews related work in wildlife drone systems and human-autonomy teaming; Section \ref{method} details our system, including the service-level objectives (SLOs), vigilance monitoring algorithm, and the human-on-the-loop (HoTL) user-interface design; Section \ref{eval} details retrospective validation methodology; we report the results in Section \ref{sec:results}; Section \ref{discussion} analyzes limitations and requirements for field deployment; and Section \ref{sec:conclusion} summarizes contributions and ongoing work.

\section{Background}
\label{background}

Effectively monitoring wildlife with drones requires balancing proximity for data quality against disturbance that compromises both animal welfare and behavioral validity \cite{hodgson2016best, duporge2021altitude, afridi2025impact}. Unlike single-object tracking, wildlife monitoring must capture multiple individuals simultaneously within their social and environmental context \cite{kline2025studying}. System performance is measured not by data volume but by the percentage of frames scientifically usable for the intended research task \cite{kline2025wildwing}.

\subsection{Automatic in-situ behavior recognition }
Recent advances enable automated classification of animal behaviors from drone footage: the KABR dataset achieved 62\% accuracy for zebra and giraffe behaviors using X3D models \cite{kholiavchenko2024kabr}, while YOLO-Behavior provides efficient classification across multiple species \cite{chan2025yolo}. The BaboonLand dataset extended capabilities to primates \cite{duporge2025baboonland}. However, these systems treat behavior recognition as a post-hoc analysis task rather than a real-time control input. Automatic estimation and prevention of fish disturbance from underwater autonomous vehicles has been proposed to detect hiding behaviors \cite{cai2025measuringminimizingdisturbancemarine}, however this approach is not readily applied to the terrestrial domain where larger mammals are more likely to flee.

\subsection{Human-autonomy teaming}
Human-on-the-loop (HoTL) systems operate autonomously by default but enable human intervention when needed \cite{wilchek2023human}. Effective HOTL design requires clear confidence indicators, graduated alerting, and mechanisms for meaningful intervention without constant attention \cite{morey2023towards}. While successful in search-and-rescue \cite{agrawal2020next} and surveillance \cite{seo_kia_2025}, few systems proactively determine when intervention is warranted based on machine learning model uncertainty.

\begin{figure*}
    \centering
    \includegraphics[width=1\textwidth]{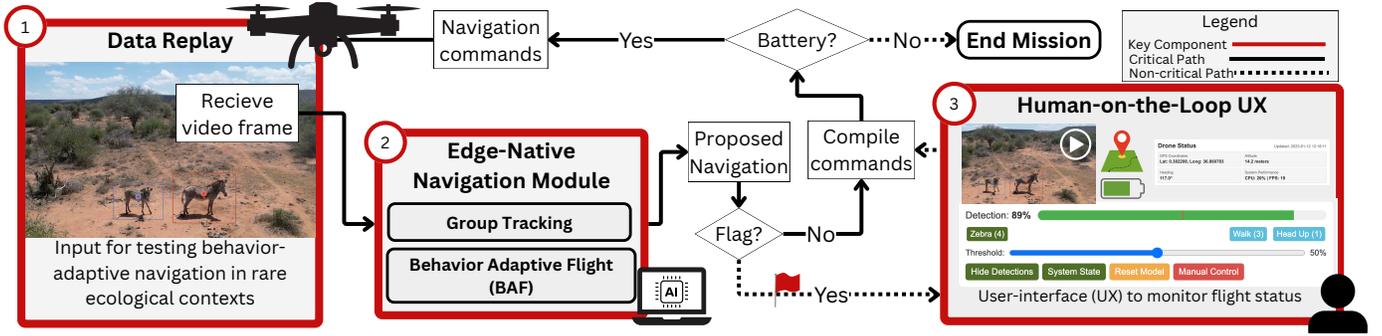}
    \caption{Edge-Native, Behavior-Adaptive Drone System Diagram. (1) Real mission replay, (2) Edge-native navigation algorithm with group-tracking and behavior-adaptive flight, (3) Human-on-the-Loop user interface.}
    \label{fig:placeholder}
\end{figure*}

\subsection{Vigilance monitoring for decision support}
Existing wildlife drone systems are implicitly human-on-the-loop but lack: (1) real-time quantification of group-level stress states, (2) graduated alerting based on behavioral leading indicators, and (3) context-tunable thresholds that account for ecological variability, such as level of human habituation, demographic factors, environmental conditions. Manual piloting cannot track individual vigilance across entire groups, while fully autonomous approaches cannot adapt to species-specific and context-specific baselines. No existing system bridges this gap by providing operators with automated vigilance monitoring as a decision-support tool. Our system integrates real-time behavior recognition with human-autonomy teaming principles to provide context-tunable vigilance monitoring. Rather than autonomous control, we focus on augmenting operator situational awareness with leading indicators that precede unrecoverable stress responses.

\section{System Design}
\label{method}

This section presents our system design and our validation approach. We establish behavioral timescale-driven service-level objectives (SLOs) (Section~\ref{sec:slos}), describe the vigilance monitoring algorithm (Section~\ref{sec:baf_nav}), and detail the operator interface (Section~\ref{sec:hotl_interface}).


\subsection{Service-Level Objectives for Behavior-Adaptive Navigation}
\label{sec:slos}

Vigilance escalation in ungulate groups follows predictable temporal dynamics \cite{kline2025kabrtools}. A study of AI-driven animal ecology studies suggests a minimum of SLO of 1 frame per second to track zebras for behavior studies \cite{kline2024characterizing}. Animals may transition behavioral states nearly instantaneously; for most drone videos shot at 30 frames per second, this latency translated to 33~ms. Analysis of adverse behavior events in zebra and giraffe missions \cite{kabr-mini-scene-videos, KABR_Raw_Videos, kline2025kabrtools, kline2025mmla} reveals a minimum 22 second warning window, the duration initial elevated vigilance, indicated by increased head-up frequency and alert postures, to full flight response, as indicated by the entire group running. To provide actionable warnings to operators, our system must detect elevated vigilance and complete classification within this window.

\textbf{Inference pipeline.} Our edge-native pipeline comprises two sequential stages:
\begin{enumerate}
    \item \textbf{Detection:} YOLOv11m~\cite{khanam2024yolov11} localizes group members for centroid-based tracking. The  latency is 4.7~ms on GPU and 183.2~ms on CPU.
    \item \textbf{Behavior classification:} YOLO-Behavior~\cite{chan2025yolo} classifies individual actions, such standing, grazing, running, with a latency of 19.1~ms on GPU and  743.7~ms on CPU.
\end{enumerate}

\textbf{Hardware implications.} GPU-accelerated edge devices (23.8~ms total) meet the 33~ms requirement. CPU-only platforms (926.9~ms) require frame sampling, thus we adopt WildWing's strategy~\cite{kline2025wildwing} of processing every 40th frame (1.33~frames per second). This approach is still able to detect behavioral changes within the critical window while maintaining lower power consumption.

\textbf{Model selection.} YOLOv11m showed superior performance for low-altitude wildlife imagery in comparative evaluation~\cite{kline2025mmla}. YOLO-Behavior reported an accuracy rate of 70\% for `Head Up' \cite{chan2025yolo}, which can be used as an indication of vigilance in ungulates \cite{kline2025kabrtools}. YOLO architectures are widely deployed in edge-based ecology applications due to proven reliability across heterogeneous hardware~\cite{kline2025studying}.

\subsection{Vigilance Monitoring Algorithm}
\label{sec:baf_nav}
Our algorithm (Algorithm~\ref{alg:nav}) aggregates individual behavioral states into a group-level vigilance metric that operators can monitor in real-time and use to inform flight decisions. To reduce false positives, we employ a 3-frame rolling window (100~ms at 30fps) 
before triggering alerts. The system only displays a warning when $S_t > \theta_S$ for 
three consecutive frames, following the strategy used in previous works \cite{kline2025kabrtools}. This prevents 
transient detections from causing alert fatigue while maintaining the 33~ms per-frame latency requirement.

\subsubsection{Group Vigilance Score}

At each time step $t$, the system processes frame $I_t$ to compute:
\begin{equation}
\label{eq:vigilance}
S_t = \frac{1}{N} \sum_{i=1}^{N} w_{l_i}
\end{equation}
where $N$ is the number of detected animals with confidence $>0.5$, $l_i$ is the classified behavior for individual $i$, and $w_{l_i}$ are behavior-specific weights. In our implementation: $w_{\text{head up}} = 1$, $w_{\text{other}} = 0$, yielding the fraction of the group exhibiting vigilance behavior. Future extensions could incorporate intermediate vigilance states (head-up frequency, ear orientation) as richer behavioral datasets become available.

\subsubsection{Context-Tunable Thresholds}

The operator defines a vigilance threshold $\theta_S$ based on ecological context:
\begin{itemize}
    \item \textbf{If $S_t \leq \theta_S$:} System displays green/yellow indicator; operator continues normal flight
    \item \textbf{If $S_t > \theta_S$:} System displays red alert with audio chime; operator decides whether to pause, retreat, or continue based on field observations
\end{itemize}

Thresholds must be tunable. Vigilance baselines vary dramatically by context: for example, in our studies habituated populations near research stations tolerate $\theta_S \approx 0.5$, while naive populations require $\theta_S \approx 0.2$. Mothers with young exhibit elevated baseline vigilance. Time of day, habitat structure, and recent disturbances all affect appropriate thresholds. No fixed policy can safely handle this variability, thus expert calibration is essential. In our retrospective analysis (Section~\ref{sec:results}), we use $\theta_S = 0.3$ as a conservative baseline, but emphasize that field deployment requires per-context calibration.

\subsubsection{Uncertainty Handling}

When YOLO-Behavior confidence drops below 0.5 for a detection, we exclude that individual from $S_t$ calculation rather than making low-confidence classifications that could trigger false alarms. If all detections fall below threshold, the system notifies the operator of degraded model performance (e.g., due to poor lighting or occlusion).

\subsection{Human-on-the-Loop Interface}
\label{sec:hotl_interface}

The HoTL interface (Figure~3, Appendix) provides situational awareness without requiring constant attention, following human-autonomy teaming principles~\cite{morey2023towards}: automation handles routine monitoring while surfacing uncertainty for human judgment. Dashboard elements include five key elements: (1) the \textbf{vigilance indicator}, a bar showing $S_t$ relative to $\theta_S$, color-coded (green $< 0.5\theta_S$, yellow $< \theta_S$, red $\geq \theta_S$); (2) a \textbf{live video} annotated with detections and behavior labels; (3) a real-time \textbf{threshold slide}, enabling adjustment of $\theta_S$ (range 0.1--0.9); (4) a \textbf{system state} live report, displaying the navigation mode, model confidence, and battery; and (5) a series of \textbf{graduated alerts}. The alerts include a visual warning as $S_t$ approaches $\theta_S$ (yellow indicator, no audio), an audio alert when $S_t > \theta_S$ (red indicator, single chime), and finally a flashing prompt if $S_t > \theta_S$ persists $>$10s. This tiered approach prevents alert fatigue while ensuring operators notice critical events~\cite{wilchek2023human}.

\section{Evaluation Methodology}
\label{eval}

\begin{figure*}
    \centering
    \includegraphics[width=1\textwidth]{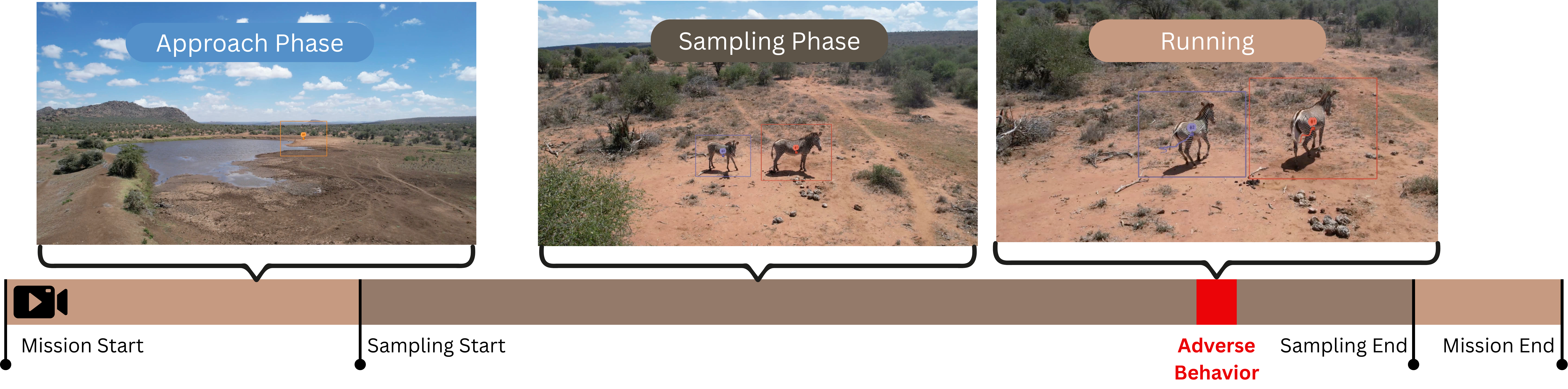}
    \caption{Experimental setup for replay - mission and sampling duration for representative video, with adverse behavior event.}
    \label{fig:placeholder}
\end{figure*}

We evaluate our system using a retrospective analysis of recorded wildlife missions from established datasets: Multi-species, multi-location, low-altitude aerial (MMLA) drone dataset \cite{kline2025mmla}, WildWing drone dataset \cite{kline2025wildwing}, and  Kenyan Animal Behavior Recognition Datasets \cite{KABR_Raw_Videos, kabr-mini-scene-videos}. Such retrospective analysis of recorded missions cannot validate closed-loop control performance, as we cannot know whether earlier intervention would have prevented escalation. However, it can establish: (1) detection accuracy on real wildlife footage, (2) inference latency compliance with behavioral timescales, (3) warning windows available to operators, and (4) the cost of reactive-only control. These capabilities are prerequisites for field deployment.

\subsection{Experimental Design}

We analyze seven wildlife monitoring missions from established datasets. For the HiTL baseline we evaluate 3 manual piloting missions from KABR~\cite{kabr-mini-scene-videos, KABR_Raw_Videos, kline2025kabrtools, kholiavchenko2024kabr}. For HoTL, we evaluate autonomous tracking without behavior adaptation from 4 missions WildWing~\cite{kline2025wildwing} and MMLA~\cite{kline2025mmla}. All missions involved 4--9 individuals in open savanna, 30~fps video, 15--25~m altitude (for demographics see Table \ref{tab:demographics}, Appendix).

\subsection{Analysis Protocol}

For each mission video, we process frames offline using detection and classification (YOLOv11m \cite{khanam2024yolov11}) and behavior labeling (YOLO-Behavior \cite{chan2025yolo}) on Apple M2 Pro chip to match field hardware. Next, we compute $S_t$ from the behavior classifications and measure \textbf{warning window}: time between first $S_t > \theta_S$ and ground-truth flight response. Afterwards, we calculate \textbf{data quality}, the  percentage of frames with confident detections and no adverse behavior ($S_t \leq \theta_S$). For our system, we simulate operator response: when $S_t > \theta_S$, assume 5-second intervention (pause/retreat), then resume when $S_t < \theta_S$ for 5 consecutive frames. This models optimistic but realistic operator response time.

\subsection{Performance Metrics}

 With only seven missions across multiple species and locations, we report descriptive statistics without significance testing. We focus on magnitude of differences and contextual interpretation (Tab. \ref{tab:metrics}), acknowledging that larger prospective studies are needed for rigorous statistical comparison.

\begin{table}[h]
\centering\small
\caption{Performance evaluation metrics}
\label{tab:metrics}
\begin{tabular}{@{}llp{3cm}@{}}
\toprule
\textbf{Category} & \textbf{Metric} & \textbf{Definition} \\ 
\midrule
\multirow{2}{*}{Detection} 
& Warning window & Time between first $S_t > \theta_S$ and flight response \\
& Detection accuracy & True positive rate for elevated vigilance \\
\midrule
\multirow{2}{*}{Data Quality} 
& Usable frames & \% frames with animals visible and $S_t \leq \theta_S$ \\
& Mission duration & Total and sampling phase time (mm:ss) \\
\midrule
Animal Welfare 
& Adverse behavior & Cumulative time where $S_t > \theta_S$ (mm:ss) \\
\midrule
\multirow{2}{*}{System} 
& Scalability & Deployment complexity (Low/Med/High) \\
& Robustness & Recovery capability (Low/Med/High) \\
\bottomrule
\end{tabular}
\end{table}

\section{Results}
\label{sec:results}

We analyzed seven wildlife monitoring missions to evaluate detection capability, quantify warning windows, and compare data quality across control approaches. Our results establish that vigilance monitoring can provide actionable alerts to operators and reveal the substantial cost of reactive-only control.

\subsection{Detection Performance and Warning Windows}

Table~\ref{tab:warning_windows} shows the critical finding: in the 4 out of 7 missions where adverse behavior occurred, our system detected elevated alert vigilance ($S_t > \theta_S$) an average of 51 seconds before full flight response. The warning window is the duration of time from the initial vigilance detection and the flight response. This warning window provides operators with actionable time to implement graduated responses, such as reducing the approach speed, pausing, or retreating, before disturbance becomes unrecoverable. The flight response is the duration of time the animals were running.

\begin{table}[h]
\centering
\small
\caption{Warning windows between alert vigilance detection and flight response}
\label{tab:warning_windows}
\begin{tabular}{@{}lccc@{}}
\toprule
\textbf{Mission} & \textbf{First Detection} &  \textbf{Alert Vigil. } & \textbf{Flight Resp.}\\
 & \textbf{(MM:SS)} & \textbf{(sec)} & \textbf{(sec)} \\
\midrule
1 & 03:45 & 53 & 19\\
2 & 03:29 & 22 & 18\\
3 & 10:20 & 38 & 51\\
4 & 01:30 & 91 & 9\\
\midrule
\textbf{Mean} & 04:46 & \textbf{51} & \textbf{24.25}   \\
\bottomrule
\end{tabular}
\end{table}

\subsection{Data Quality Comparison}
\begin{table*}[t]
\centering
\renewcommand{\arraystretch}{1.3}
\caption{Summary of data quality and system performance for data collection methods}
\label{tab:methodology_comparison}
\begin{tabular}{|l|c|c|c|c|c|}
\hline
\multirow{2}{*}{\textbf{Data Collection Method}} & \multicolumn{3}{c|}{\textbf{Data Yield \& Volume}} & \multicolumn{2}{c|}{\textbf{System Performance}} \\
\cline{2-6}
 & \textbf{Usable Frames (\%)} & \textbf{Adverse Behavior} & \textbf{Mission Time} & \textbf{Scalability} & \textbf{Robustness} \\
 & Total / Sampling & (MM:SS) & Total / Sampling &  &  \\
\hline
HITL & 
 71.9\% / 94.8\% & 
 00:14 & 
 \textbf{11:58 / 9:06} & 
\cellcolor{red!25} Low & 
\cellcolor{yellow!25} Medium \\
\hline
HOTL & 
 81.4\% / \textbf{98.2\%} & 
 00:02 & 
 04:18 / 2:03 & 
\cellcolor{green!25} \textbf{High} & 
\cellcolor{green!25} \textbf{High} \\
\hline
BAF & 
 \textbf{86.8\%} / 
95.2\% & 
 \textbf{00:01} & 
 04:42 / 4:09 & 
\cellcolor{yellow!25} Medium & 
\cellcolor{red!25} Low \\
\hline
BAF + HOTL & 
 82.8\% / 97.0\% & 
\textbf{00:01} & 
 07:35 / 5:04 & 
\cellcolor{green!25} \textbf{High} & 
\cellcolor{green!25} \textbf{High} \\
\hline
\end{tabular}
\vspace{2mm}

\footnotesize{\textit{Note:} See Table~\ref{tab:metrics} for evaluation metric details. Color-coded performance indicators: \colorbox{green!25}{green} = good, \colorbox{yellow!25}{yellow} = medium, \colorbox{red!25}{red} = poor. \\ \textbf{Bold} values indicate best performance in each category.}
\end{table*}

Table~\ref{tab:methodology_comparison} presents comprehensive performance across all four approaches.
 The manual piloting (HiTL) baseline control resulted in 71.9\% usable frames with 14 seconds average adverse behavior duration per mission. While achieving longest total mission times (11:58 average), the high disturbance rate and inconsistent data quality reflect the attentional bottleneck problem—operators cannot simultaneously control flight and monitor individual vigilance states.
The HoTL without behavior adaptation, i.e. autonomous tracking without vigilance monitoring proposed in \cite{kline2023framework}, achieved 81.4\% usable frames with minimal adverse behavior (0.3s average), but at the cost of substantially shorter missions (4:18 total, 2:03 sampling). This suggests conservative flight strategies that maintain distance but sacrifice data resolution and coverage.

To determine projected system performance, a simulated 5-second operator intervention when $S_t > \theta_S$ yields 82.8\% usable frames with 1-second adverse behavior duration, while maintaining 7:35 mission duration (63\% of HiTL baseline). This represents a 93\% reduction in adverse behavior compared to manual piloting, while preserving 115\% of data volume compared to HoTL-only approaches.

\subsection{System Performance Assessment}


Manual piloting requires continuous operator attention per drone, limiting deployment scale. HoTL systems reduce cognitive load through autonomous tracking, but lack behavioral awareness as operators must still monitor video streams for signs of distress. Our system addresses this by automating vigilance monitoring: the system tracks individual behaviors across the group and surfaces only critical events (red alerts when $S_t > \theta_S$), allowing operators to supervise rather than continuously control the system.



Recovery capability varies fundamentally across approaches. Manual piloting and HoTL systems can adapt to unexpected events, while autonomous behavior adaptive approaches must terminate missions when thresholds are breached as there is no reliable autonomous recovery mechanism. Our system combines strengths: automated monitoring detects issues early, then human judgment determines appropriate recovery, whether to pause briefly (4.3s average in our simulations), retreat and reapproach from different angle, or terminate if animals remain stressed. This hybrid approach proved more robust than either pure autonomy or pure manual control in our retrospective analysis.

We note these results represent detection capability, not closed-loop control validation. We cannot determine from historical data whether preemptive intervention would have prevented escalation, only that warning signals existed in time for response. 
Future field trials with live closed-loop control are necessary to validate actual prevention capability.
\section{Discussion}
\label{discussion}

Our analysis establishes three critical capabilities. First, sufficient warning windows exist. There was an average 51-second warning between elevated vigilance and flight response provides actionable time for operator intervention. Two, edge hardware meets latency requirements. The proposed edge-native navigation module has a 23.8~ms inference latency, enabling real-time vigilance monitoring within behavioral timescales which is 33~ms for live video monitoring. Third, reactive control is costly. Manual piloting results in 93\% more adverse behavior and 18\% lower data quality compared to vigilance-informed approaches, justifying the complexity of behavior-adaptive systems.

Next steps include closing the control loop, moving from decision support to adaptive control.
Our current system implementation provides operators with vigilance state information but does not directly modulate flight commands. Future work will integrate vigilance scores as control inputs, implementing graduated responses where approach velocity scales inversely with herd vigilance levels. This would enable feedforward control, adaptive speed profiles, and shared autonomy. For feedforward control, the approach speed is slowed when elevated vigilance is detected, rather than waiting for operator intervention. The system will dynamically adjust flight parameters based on real-time behavioral feedback for adaptive speed profiles. The system proposes graduated responses (slow/pause/retreat) while the operator retains override authority.

However, closing this loop introduces safety-critical requirements that retrospective analysis cannot validate. Field trials with live closed-loop control are necessary to evaluate: (1) whether preemptive speed adjustments actually prevent escalation or simply delay it, (2) how different species respond to varying approach dynamics, and (3) whether the system can safely recover from false positives without mission failure. 

\section{Conclusion}
\label{sec:conclusion}

Wildlife monitoring drones face a fundamental challenge: operators must track individual vigilance states across entire herds while simultaneously controlling flight operations. This attentional bottleneck leads to reactive control that results in 14 seconds average adverse behavior per mission and 71.9\% usable data yield. We present our system, a configurable decision-support system that addresses this limitation through automated herd-level vigilance monitoring with context-tunable thresholds.
Our retrospective analysis of seven wildlife monitoring missions establishes three key capabilities. First, sufficient warning windows exist: analysis reveals warning windows of an average of 51s between initial 
vigilance detection and flight response, demonstrating that real-time monitoring  provides actionable intervention opportunities. Second, edge hardware meets behavioral timescale requirements: our 23.8~ms inference pipeline operates well within the 33~ms SLO. Third, reactive control is costly: manual piloting produces 93\% more adverse behavior than vigilance-informed approaches, justifying the system complexity.


Wildlife monitoring exemplifies a broader challenge in edge AI. Deploying intelligent systems in domains where context varies unpredictably have consequences to animal welfare. Our approach demonstrates that exposing decision thresholds to domain experts, rather than pursuing full automation, can bridge the gap between AI capabilities and ecological requirements. This human-AI partnership model may generalize to other edge applications where operational context cannot be fully specified in advance.
%
Our proposed system provides decision support but does not yet close the control loop. Future work will integrate vigilance scores as feedforward control signals, implementing graduated responses where approach velocity scales inversely with group stress levels. Future work could conduct validation trials to determine whether preemptive speed adjustments prevent escalation. Further investigation in partnership with ecologists could determine how different species respond to varying approach dynamics, and whether the system safely recovers from false positives. 
%


{\small
\vspace{4pt} \noindent {\bf Acknowledgments:} 
This project is supported by the AI Institute for Intelligent Cyberinfrastructure with Computational Learning in the Environment (ICICLE), which is funded by the US National Science Foundation under Award \#2112606. 
}

\section{Appendix}

\subsection{Algorithmic Details and Mathematical Notation}

\begin{algorithm}[h]
\footnotesize
\caption{HoTL-BAF: Vigilance Monitoring for Operator Decision Support}
\label{alg:nav}
\begin{algorithmic}[1]
\Require Video stream $I$, detection model $\mathcal{C}_{\text{detect}}$, behavior model $\mathcal{C}_{\text{behav}}$, vigilance threshold $\theta_S$, confidence threshold $\theta_c = 0.5$
\Ensure Real-time vigilance score $S_t$ and operator alerts

\While{mission active}
    \State $I_t \leftarrow$ acquireFrame()
    \Statex
    \Statex \textit{// Detection and Tracking}
    \State $\mathcal{B}_t \leftarrow \mathcal{C}_{\text{detect}}(I_t)$ \Comment{Detect animals}
    \State $\mathcal{B}_t^{\text{conf}} \leftarrow \{(b_i, p_i) \in \mathcal{B}_t : p_i > \theta_c\}$ \Comment{Filter low confidence}
    
    \If{$\mathcal{B}_t^{\text{conf}} = \emptyset$}
        \State updateGUI(NO\_DETECTIONS)
        \State \textbf{continue}
    \EndIf
    \Statex
    \Statex \textit{// Behavior Classification}
    \State $N \leftarrow |\mathcal{B}_t^{\text{conf}}|$
    \State $n_{\text{adverse}} \leftarrow 0$
    
    \For{each $(b_i, p_i) \in \mathcal{B}_t^{\text{conf}}$}
        \State $(\ell_i, q_i) \leftarrow \mathcal{C}_{\text{behav}}(I_t, b_i)$ \Comment{Classify behavior}
        \If{$q_i > \theta_c$ \textbf{and} $\ell_i = \text{vigilant}$}
            \State $n_{\text{adverse}} \leftarrow n_{\text{adverse}} + 1$
        \EndIf
    \EndFor
    \Statex
    \Statex \textit{// Vigilance Scoring}
    \State $S_t \leftarrow \frac{n_{\text{adverse}}}{N}$ \Comment{Fraction of herd vigilant}
    \Statex
    \Statex \textit{// Operator Alerts}
    \If{$S_t \geq \theta_S$}
        \State displayAlert(RED, AUDIO\_CHIME)
        \State updateGUI($S_t$, $\mathcal{B}_t^{\text{conf}}$, HIGH\_VIGILANCE)
    \ElsIf{$S_t \geq 0.5 \theta_S$}
        \State displayAlert(YELLOW, NO\_AUDIO)
        \State updateGUI($S_t$, $\mathcal{B}_t^{\text{conf}}$, ELEVATED)
    \Else
        \State displayAlert(GREEN, NO\_AUDIO)
        \State updateGUI($S_t$, $\mathcal{B}_t^{\text{conf}}$, NORMAL)
    \EndIf
\EndWhile
\end{algorithmic}
\end{algorithm}

\begin{table}[h]
\centering\small
\caption{Mathematical notation and definitions}
\label{tab:notation}
\begin{tabular}{@{}cl@{}}
\toprule
\textbf{Symbol} & \textbf{Definition} \\
\midrule
\multicolumn{2}{@{}l}{\textit{Time and Frames}} \\
$t$ & Time step / frame index \\
$I_t$ & Video frame at time $t$ \\
\midrule
\multicolumn{2}{@{}l}{\textit{Detection and Tracking}} \\
$\mathcal{B}_t$ & Set of detected bounding boxes at time $t$ \\
$b_i$ & Bounding box for individual $i$ \\
$p_i$ & Detection confidence for individual $i$ \\
$N$ & Number of detected animals \\
$\mathbf{c}_t$ & Herd centroid position at time $t$ \\
\midrule
\multicolumn{2}{@{}l}{\textit{Behavior Classification}} \\
$\mathcal{C}_{\text{detect}}$ & Detection model (YOLOv11m) \\
$\mathcal{C}_{\text{behav}}$ & Behavior classification model (YOLO-Behaviour) \\
$\ell_i$ & Behavior label for individual $i$ \\
& \quad $\in \{\text{standing, grazing, running, ...}\}$ \\
$q_i$ & Classification confidence for individual $i$ \\
$w_{\ell_i}$ & Behavior-specific weight ($w_{\text{running}} = 1$, $w_{\text{other}} = 0$) \\
\midrule
\multicolumn{2}{@{}l}{\textit{Vigilance Scoring}} \\
$S_t$ & Herd vigilance score at time $t$ (Equation~\ref{eq:vigilance}) \\
$\theta_S$ & Vigilance threshold (user-adjustable, default 0.3) \\
$\theta_c$ & Confidence threshold for detections (default 0.5) \\
$n_{\text{adverse}}$ & Number of individuals exhibiting adverse behavior \\
\midrule
\multicolumn{2}{@{}l}{\textit{Performance Metrics}} \\
$\tau_d$ & Detection latency (warning window duration) \\
$\tau_h$ & Hover response time \\
\bottomrule
\end{tabular}
\end{table}

\subsection{Drone mission details.}
See Table \ref{tab:demographics}.


\begin{table*}
\centering
\begin{threeparttable}
\caption{Summary of location, data collection technique, herd size and species for each drone mission used in evaluation.}
\setlength{\tabcolsep}{0.35em}
\renewcommand{\arraystretch}{1.2}
\begin{tabular}{|c|c|c|c|c|p{5cm}|}
\hline
\textbf{Session} & \textbf{Collection Location} & \textbf{Collection Technique} & \textbf{Source Dataset} & \textbf{Herd Size} & \textbf{Species} \\
\hline
1 & Mpala, Kenya\tnote{1} & Manual & KABR\tnote{4} & 4 & 2 plains zebras (female), 1 adult female Grevy's zebra with foal, gazelles captured in background. Video recorded as herd leaving watering hole. \\
\hline
2 & Mpala, Kenya\tnote{1} & Manual & KABR\tnote{4} & 8 & Giraffes in bushy area. \\
\hline
3 & Mpala, Kenya\tnote{1} & Manual & KABR\tnote{4} & 5 & Plains zebras. \\
\hline
4 & The Wilds, USA\tnote{3} & HOTL\tnote{6} & WildWing\tnote{5} & 5 & Grevy's zebras \\
\hline
5 & The Wilds, USA\tnote{3} & HOTL\tnote{6} & WildWing\tnote{5} & 2 & 5 giraffes total in pasture \\
\hline
6 & The Wilds, USA\tnote{3} & HOTL\tnote{6} & WildWing\tnote{5} & 8 & Przewalski's horses\\
\hline
7 & Ol Pejeta Conservancy, Kenya\tnote{2} & HOTL\tnote{6} & MMLA\tnote{7} & 9 & Plains zebras \\
\hline
\end{tabular}

\begin{tablenotes}[flushleft]
\footnotesize
\item[1] \textbf{Mpala Research Centre} is located in Laikipia County, Kenya~\cite{mpala}. 
KABR datasets were collected here using a \textit{DJI Mavic Air~2} drone operated manually to capture fine-grained animal behaviors~\cite{KABR_Raw_Videos,kabr-mini-scene-videos}.
\item[2] \textbf{Ol Pejeta Conservancy}, also in Laikipia County, Kenya~\cite{olpejeta}, hosted a portion of the MMLA data collections using the \textit{WildWing} semi-autonomous aerial system~\cite{kline2025wildwing}, based on \textit{Parrot Anafi} drones operating in Human-on-the-Loop (HOTL) mode.
\item[3] \textbf{The Wilds} is a 10{,}000-acre conservation center in Cumberland, Ohio, USA~\cite{wilds}, managed by the Columbus Zoo and Aquarium. Data collection also employed the \textit{WildWing} semi-autonomous tracking system~\cite{kline2025wildwing} with \textit{Parrot Anafi} drones for long-duration behavioral observations of Grevy’s zebras, giraffes, and Przewalski’s horses.
\item[4] \textbf{KABR} – Kenyan Animal Behavior Recognition dataset, comprising manually flown aerial videos of zebras and giraffes for behavior classification and tracking~\cite{KABR_Raw_Videos,kabr-mini-scene-videos, kholiavchenko2024kabr, kline2025kabrtools}.
\item[5] \textbf{WildWing} – Open-source, semi-autonomous drone system for adaptive wildlife monitoring~\cite{kline2025wildwing}.
\item[6] \textbf{HOTL (Human-on-the-Loop)} – Semi-autonomous operational mode where a human operator supervises flight and may intervene if necessary.
\item[7] \textbf{MMLA} – Multi-Environment, Multi-Species, Low-Altitude drone dataset \cite{kline2025mmla}. This work used the portion of the dataset collected at Ol Pejeta Conservancy (OPC) \cite{mmla_opc}.
\end{tablenotes}

\end{threeparttable}
\label{tab:demographics}
\end{table*}

\subsection{User Interface Screenshot}
\begin{figure*}
    \centering
    \includegraphics[width=0.8\linewidth]{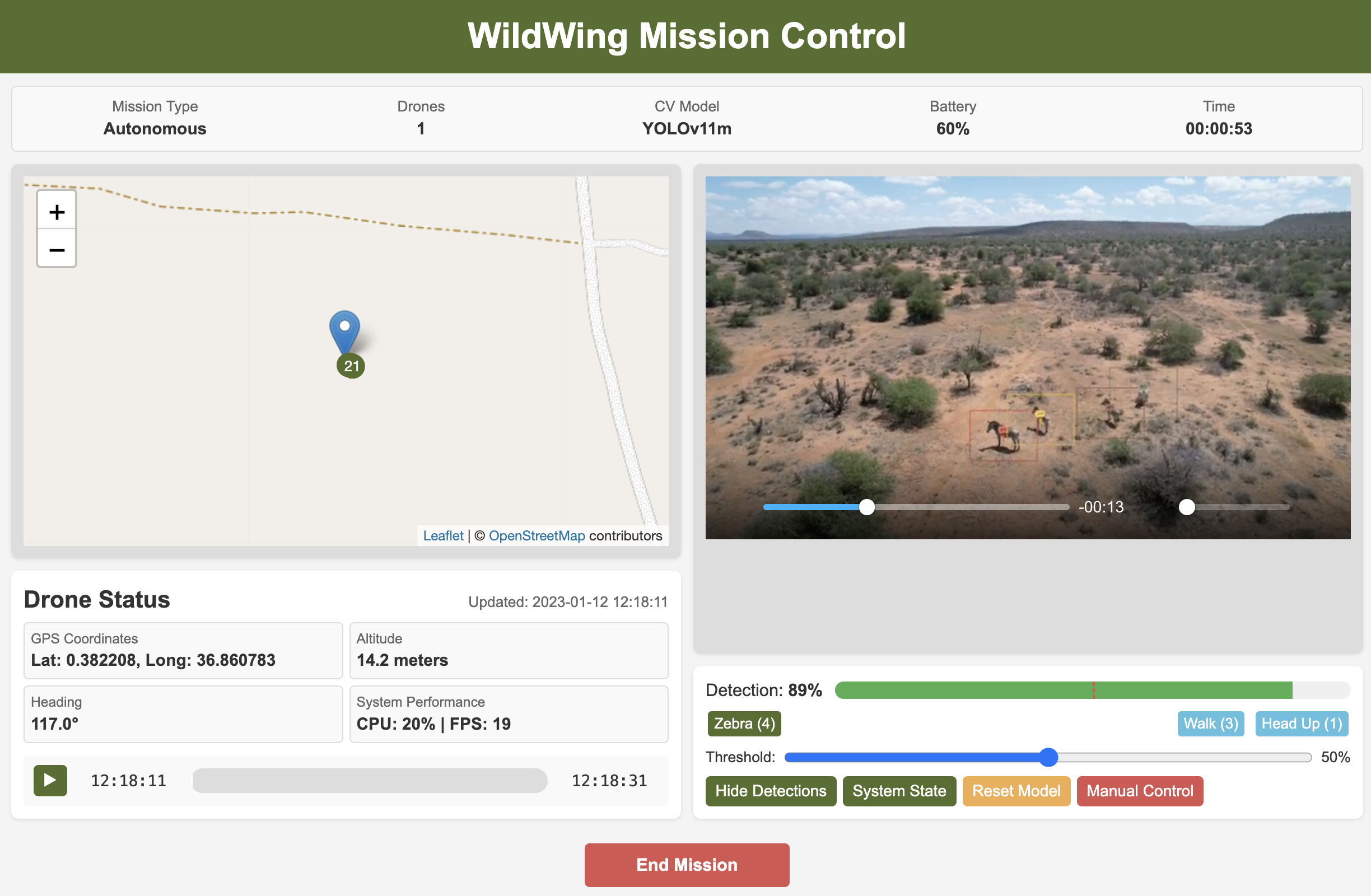}
    \caption{Human-on-the-Loop User Interface for Behavior-Adaptive Drone System Monitoring and Vigilance Monitoring}
    \label{fig:ux}
\end{figure*}

See Figure \ref{fig:ux}.

\newpage

{\small
\bibliographystyle{abbrv}
\bibliography{ref}
}

\end{document}